\documentclass[11pt]{article}
\usepackage[utf8]{inputenc}
\usepackage[left, pagewise]{lineno}

\usepackage{xcolor}
\usepackage[T1]{fontenc}
\usepackage{graphicx}
\usepackage[export]{adjustbox}

\usepackage{amsmath}

\usepackage[scr=rsfs]{mathalpha}
\usepackage{amsfonts}
\usepackage{amssymb}
\usepackage[version=4]{mhchem}
\usepackage{stmaryrd}
\usepackage[hidelinks]{hyperref}
\usepackage{subcaption}
\usepackage[numbers,square]{natbib}

\usepackage{authblk}

\usepackage{stmaryrd}
\usepackage{comment}
\usepackage{soul}

\usepackage[ruled,vlined]{algorithm2e}
\usepackage{float}
\usepackage{xcolor}

\newcommand{\la}[1]{\hat{#1}} 

\title{Solving the six-dimensional Vlasov--Maxwell System with Active Flux and Splitting Methods}

\author[1]{G. Gr\"unwald}
\author[2]{L. Hensel}
\author[1,2]{M. Deisenhofer}
\author[1,2]{S. Lautenbach}
\author[2]{K. Kormann}
\author[1]{R. Grauer}

\affil[1]{Theoretical Physics I, Ruhr-Universit\"at Bochum, Universit\"atsstra{\ss}e 150, D-44801 Bochum, Germany}
\affil[2]{Numerical Mathematics, Ruhr-Universit\"at Bochum, Universit\"atsstra{\ss}e 150, D-44801 Bochum, Germany}
\date{}
\begin{document}
	\maketitle
	
	\begin{abstract}
	Active Flux (AF) is a modified Finite Volume method that evolves additional Degrees of Freedom (DoF) located on the cell interfaces to compute high-order approximations to the numerical fluxes through the respective interface. 
	We present an AF-based scheme for the simulation of collisionless plasmas described by the Vlasov equation coupled with Maxwell's equations. In order to limit the DoF in high dimensional settings we employ operator splitting. The resulting one-dimensional advection equations can be solved efficiently and with low implementation complexity, making it a very fast alternative to standard Finite Volume methods. We compare our scheme’s performance with a related Finite Volume method based on the semi-Lagrangian approach. We find that, as a consequence of its compact stencil, the AF scheme has significantly lower dissipation and reduced anisotropy, and thus produces results on par with or even superior to the benchmark for standard test cases reproducing important kinetic phenomena, while also offering lower computational cost.
\end{abstract}

\section{Introduction}
The kinetics of collisionless plasmas, such as those occurring between the Sun and Earth, are described by the Vlasov equation, which forms a self-consistent system with Maxwell’s equations. Due to its high dimensionality arising from three spatial directions, the corresponding velocities, and the time, the simulation of the Vlasov--Maxwell system is inherently costly. For this reason, it can only be afforded in limited regions. To nevertheless capture the entirety of large-scale physical phenomena such as jets caused by magnetic reconnection, kinetically treated areas wherever non-equilibrium dynamics are particularly important to recognize can be coupled to two-fluid, and finally MHD simulation \cite{Allmannrahn2024,lautenbach-grauer:2018,lautenbach_2024_10547265}.

The Vlasov equation is commonly solved using semi-Lagrangian methods. They make use of the advective form of the Vlasov equation by tracing the solution backward in time along the characteristic equations to update it at fixed grid points. To find the value of the solution at the origin of the characteristic curve, interpolation to the desired order is used \cite{staniforth-cote:1991}. Semi-Lagrangian methods as such do not conserve mass. This shortcoming was prominently remedied by the introduction of the Positive and Flux-Conservative (PFC) method \cite{Filbet2001}, where fluxes through the cell interfaces are calculated using semi-Lagrangian considerations: they are the spatial integral of the interpolated solution between the respective cell border and the origin of the characteristic ending at the adjacent cell border. Depending on the reconstruction and its stencil, higher-order accuracy can be reached. The ingoing and outgoing fluxes of a cell are then balanced to update the cell average of the solution like it is done in Finite Volume (FV) methods, making PFC mass conservative. PFC is known to be stabilized by its numerical dissipation, as already noted in the original paper. It also uses slope limiters to prevent oscillations at sharp gradients from producing negative densities and to respect the maximum principle.

Discontinuous Galerkin methods \cite{rossmanith-seal:2011,Juno2018} achieve higher order with a localized stencil and are also conservative. Due to the discontinuity of the reconstruction, storage requirements are however decreased. Oftentimes, a multi-stage time evolution such as the Runge-Kutta method is used, which requires frequent boundary exchanges and thereby increases communication expense.

Efforts have been made to also conserve higher moments of the Vlasov equation such as the total momentum and energy (e.g. \cite{Allmannrahn2022a, Juno2018}). 

An important approach to reduce cost is operator splitting, where the problem is broken down into a series of lower-dimensional subproblems. For the Vlasov--Poisson system, this dates back to \cite{Cheng1976}.
In addition, low-rank tensor methods \cite{Kormann2015,EiLu2018,Allmannrahn2022b,EiKoKu2025} have shown promising results.
An alternative to grid-based methods are Particle-In-Cell (PIC) methods \cite{Birdsall}, where the frame of reference moves along the characteristic trajectories of a finite number of (super) particles.
In this context, the preservation of inherent invariants such as Gauss' law by exploiting the Hamiltonian structure of the Vlasov--Maxwell system has been discussed in \cite{SqQiTa2012,Kraus2017,CamposPinto2022,Kormann2024}. The electromagnetic fields are herein discretized using either mimetic finite-difference or finite-element methods. The second-order Finite-Difference Time-Domain method introduced by Yee \cite{Yee1966} can be derived as a special case of these methods and shares their structure-preserving properties. Notably, energy conservation in addition to the preservation of Gauss' law was achieved in \cite{Kormann2021} by means of Hamiltonian splitting. 
PIC and grid-based Vlasov codes are both widely employed in their respective domains: Vlasov methods are preferred when high accuracy in velocity space and low noise are essential, whereas PIC methods are favored for large-scale three-dimensional simulations and for cases involving particle beams.

In this work we propose the use of the Active Flux (AF) method in combination with operator splitting and the FDTD method to efficiently solve the Vlasov--Maxwell system. The first analysis of AF as a third-order conservative difference method for one-dimensional linear advection was given by van Leer \cite{van1977towards}, while the idea has later been extended to more general hyperbolic conservation laws by Eymann and Roe \cite{Roe2011, Roe2013}. AF is a Finite Volume method, where the interface fluxes are constructed from additional point Degrees of Freedom (DoF) located at the cell interfaces. These can be evolved in time in any given way, which can be non-conservative. For the Vlasov equation, it is convenient to approach this the semi-Lagrangian way. This leads to a compact temporal stencil. Through the additional DoF per cell, the method achieves third-order accuracy while keeping a compact stencil in space. This is in contrast to classical Finite Volume or semi-Lagrangian methods. In fact, by adding DoF to the inside of the cell, Active Flux schemes of arbitrary order of accuracy can be designed \cite{abgrall-etal:2023b}. In Active Flux methods, the interface DoF are shared between neighboring cells, resulting in a continuous reconstruction and reduced storage overheads compared to standard DG methods, particularly for many dimensions due to the increased surface-to-volume ratio. See \cite{Roe_DG} for a direct comparison of the two methods.

The compact stencil keeps the domain of dependence physically meaningful and keeps the communication frequency low. Numerically speaking, this relaxes the time-step restriction, and reduces anisotropy and dissipation. Due to the additional sampling points, high-frequency effects can be resolved more effectively than by standard Finite Volume methods. These properties make the AF method desirable for the simulation of the Vlasov equation, as only a very limited number of grid cells can here be afforded. 

Over the last years, Active Flux has gained a lot of interest mainly
in the fluid community and has been taken from simplices to Cartesian grids \cite{barsukov-holm-etal:2019,barsukow:2021,chudzik-helzel-kerkmann:2021,bai-roe:2021}.
In \cite{kiechle2023active}, an Active Flux method without dimensional splitting was proposed for the 1D1V Vlasov--Poisson system and a limiter introduced in \cite{kiechleNew}.
In fact, Active Flux methods were designed to be truly multidimensional. For reasons we have laid out in \cite{Hensel2024}, and which we will reiterate in Sec.~\ref{sec:AF}, this becomes however unfeasible in higher dimensions leading up to 6D, if a straightforward tensorial approach as it is custom in Finite Elements is used. Instead, we use operator splitting to break the problem down into low dimensional problems, allowing us to apply AF in one dimension and therefore drastically reducing the number of DoF involved in a given update substep. This makes it possible to exploit the advantages of AF which we summarize here: The stencil being compact in space (as opposed to other third-order Finite Volume methods) as well as in time (allowing for a maximized time-step, as opposed to e.g. Galerkin schemes), and the continuous reconstruction reducing memory overhead compared to discontinuous methods.
In \cite{Hensel2024}, three split-step AF methods were proposed and tested for the 1D1V Vlasov--Poisson system. The methods were designed such that their extension to the full 6D case would be feasible and straightforward. Two of these methods, one using numerical fluxes explicitly integrated over the cell interface, the other one based on the discrepancy distribution formulation of AF \cite{he2020treatment}, were shown to reach third order in space. They were, however, more costly with worse scaling properties than the remaining method, which we will apply to the Vlasov--Maxwell system up to 6D in this work. While it converged to second order, it is important to note that in high-dimensional Vlasov simulations, the resolution, in particular in velocity space, often needs to be such that the error behavior of the method is not in asymptotic regime. Even though the three presented methods had differing convergence behavior, they performed equally well with regards to the reproduction of small-scale structures in phase-space and the damping / growth rate in the standard electrostatic test-cases of Landau damping and two-stream instability. This is possibly owed to their common third-order yet compact reconstruction of the solution. Following this observation, we now aim to obtain useful results at feasible computing cost, oftentimes not investing into the formal order of accuracy. This manifests in our choice of the classical Yee-scheme, as well as linear interpolations. This is in alignment with the common practice of combining higher-order approximations as part of the Vlasov solver to capture the filamentation or phase-mixing developing in phase space (or conversely apply a filter to smooth it away) with a less accurate but economic Maxwell solver, as here the solution tends to be smoother and more forgiving \cite{Umeda2009}. Similarly, we use the second-order Strang splitting. Despite efforts being made to reduce the splitting error of the rotational term (e.g. \cite{Schmitz2006b,Bernier2020,Schild2025}), Strang splitting is commonly used also for the fully electromagnetic problem, as it still yields accurate enough results while being more affordable \cite{Valentini2007,Palmroth2018}.

\section{The Vlasov--Maxwell System}

Collisionless plasmas are governed by the Vlasov equation which describes the evolution of the distribution function $f_s(\mathbf x, \mathbf v, t)$: 
\begin{equation}\label{eq:vlasov}
\partial_t f_s +\mathbf{v} \cdot \nabla_x f_s + \frac{q_s}{m_s} (\mathbf E + \mathbf v \times \mathbf B) \cdot \nabla_v f_s = 0.
\end{equation}

The magnetic and electric fields $\mathbf B$ and $\mathbf E$ are determined by Maxwell's equations 
\begin{align}\label{eq:maxwell} 
\partial_t \mathbf B & = -\nabla \times \mathbf E\,,\quad
\partial_t \mathbf E = c^2 \left(\nabla \times \mathbf B - \mu_0 \mathbf j\right)\\
\nabla \cdot \mathbf B & = 0\,,\qquad \quad
\nabla \cdot \mathbf E = \frac{\rho}{\varepsilon_0}\,,
\end{align} where we refer to the last constraint as Gauss' law.
Here, $\rho$ denotes the charge density, and $\mathbf j$ is the current density, which couples Maxwell's equation to the Vlasov equation. Since the electromagnetic fields are determined by the distribution function, but are used to evolve the very same, they are often called self-consistent.
Constants are the vacuum speed of light $c$, the permeability $\mu_0$ and the permittivity $\varepsilon_0$. In the electrostatic limit, where $\mathbf{B} = 0$ and $c \to\infty$ ,  $\mathbf{E}$ is described by Poisson's equation
\begin{align}\label{eqn:Poisson}
\Delta \phi & = -\frac{\rho}{\epsilon_0} \; , \;\; \mathbf{E} = - \nabla \phi \; ,
\end{align}
which is now coupled to the Vlasov equation only via the charge density $\rho$.
The computation of the source terms 
\begin{align}\label{eqn:rho_j}
\rho := \sum_s q_s n_s, \quad
\mathbf{j} := \sum_s q_s n_s \mathbf{u}_s
\end{align}
requires the zeroth and first moments of $f_s$, respectively the particle density $n_s$ and the bulk velocity $\mathbf{u}_s$: 
\begin{align}
n_s(\mathbf{x},t) & := \int_{\mathbb{R}^d} f_s(\mathbf{x},\mathbf{v},t) \, d\mathbf{v} \\
\mathbf{u}_s (\mathbf{x},t) & := \frac{1}{n_s(\mathbf{x},t)}\int_{\mathbb{R}^d}
\mathbf{v} f_s(\mathbf{x},\mathbf{v},t) \,
d\mathbf{v}  
\end{align}

\section{Numerical Setup}
\subsection{Operator Splitting}
Writing the Vlasov equation as
\begin{align}
\partial_t f_s+(\mathscr{A}+\mathscr{B}) f_s=0
\end{align}
with the operators
\begin{align}
\mathscr{A}&=\mathbf{v} \cdot \nabla_{\mathbf{x}}\\
\mathscr{B}&=\frac{q_s}{m_s}(\mathbf{E}+\mathbf{v} \times \mathbf{B}) \cdot \nabla_{\mathbf{v}},
\end{align}
we can split it into  the two advection equations
\begin{align}
\partial_t f_s+\mathscr{A} f_s&=\partial_t f_s+\mathbf{v} \cdot \nabla_{\mathbf{x}} f_s=0 \\
\partial_t f_s+\mathscr{B} f_s&=\partial_t f_s+\frac{q_s}{m_s}(\mathbf{E}+\mathbf{v} \times \mathbf{B}) \cdot \nabla_{\mathbf{v}} f_s=0,
\end{align}
accepting the introduction of a splitting error we discuss in this section.

We can write the time evolution of a system
$\partial_t f+\mathscr{S} f=0, \quad f(t=0)=f_0$ as $f(\Delta t)=\exp (\Delta t \mathscr{S}) f_0=: T(\mathscr{S}) f_0$, if $\mathscr{S}$ is time-independent. If $\mathscr{S}$ is composite, we can split this into $f(\Delta t)=T(\mathscr{S}_1) T(\mathscr{S}_2) f_0$, and it is exact if $\mathscr{S}_1$ and $\mathscr{S}_2$ commute, according to the Baker-Campbell-Hausdorff formula, and correct to first order otherwise. This type of splitting is then known as Lie splitting.

As $\mathscr{A}$ and $\mathscr{B}$ do not commute, we instead use the second order correct Strang splitting (first applied to the Vlasov equation in \cite{Cheng1976}). Here a full time step of one of the operators is sandwiched into two half-timesteps of the other, with the role assignment being arbitrary:
\begin{align} & f(t+\Delta t)=T(\mathscr{S}_1+\mathscr{S}_2) f(t)=T\left(\frac{1}{2} \mathscr{S}_1\right) T(\mathscr{S}_2) T\left(\frac{1}{2} \mathscr{S}_1\right) f(t) + \mathscr{O}\left(\Delta t^3\right) .
\end{align}
Higher-order splittings can be constructed (see eg. \cite{yoshida1990construction}). 

To further reduce dimensionality, we split the sub-systems into their parts. In the case of $\mathscr{A}$,can be split without error as the $(v\nabla_x)_i$ commute. The components of $\mathscr{B}$,  $((\mathbf{E} + \mathbf{v}\times \mathbf{B}) \cdot \nabla_{\mathbf{v}})_i$, on the other hand do not commute due to the cross-product and we again employ Strang splitting. We note that the use of Strang splitting adds anisotropy to the scheme. An alternative without this flaw would be the backsubstitution method \cite{Schmitz2006b} based on the cascade interpolation \cite{leslie-purser:1995}.  Conveniently, the advection velocities of the 6 one-dimensional advection equations we are now left with, are constant in the respective spaces. The E-M fields however are time-dependent, and the evolution of such operators is $ T(\mathscr{S}) = \exp (\int dt \mathscr{S})$. To second order, this can be approximated using the midpoint in time. When considering the Vlasov--Poisson system, this can easily be realized by evaluating $\mathbf{E}$ between a first half-timestep in configuration space and the following full time-step in velocity space \cite{Cheng1976}. This gives $\mathbf{E}^{t+1/2}$ as the charge density does not change during the velocity update. For the Vlasov--Maxwell system, where the source term $\mathbf{j}$ now weights $f$ with $\mathbf{v}$, the calculation of $\mathbf{E}^{t+1/2}$ would become implicit in this approach.  
We instead choose a time-stepping sequence as described in \autoref{alg:timestepping}. Initially, half a Maxwell update is performed using $\mathbf{j}^{t=0}$ to approximate the fields at $t+1/2$, i.e. an Euler step. The magnetic field is therein update to full time, in preparation of the staggered time grid of the FDTD solver used to evolve Maxwell's equations. The Maxwell update is now leapfrogged with the Vlasov update.
This can be understood as Strang splitting of the Vlasov- from the Maxwell part, with two consecutive Maxwell half-steps combined into one inside the time step loop. The roles of $\mathbf{x}$ and $\mathbf{v}$ in the Strang sequence are arbitrary to second order. 

The execution of the individual steps is explained in the following sections.
\SetKwBlock{Timestep}{Time step}{end}
\begin{algorithm}
\caption{Time stepping of the Vlasov--Maxwell schemes}\label{alg:timestepping}
\textbf{Initialize }

\smallskip
Half Maxwell update:

\quad Advance $\mathbf{E}^{t=0}$ to $\mathbf{E}^{t+1/2}$ using $\mathbf{j}^{t=0}$

\quad Advance $\mathbf{B}^{t=0}$ to $\mathbf{B}^{t+1}$, interpolate to $t+1/2$

\Timestep{
	Interpolate fields to center of Yee-cube
	
	\textbf{Full Vlasov update}
	
	\quad Half velocity step using $\mathbf{E}^{t+1/2}$, $\mathbf{B}^{t+1/2}$:
	
	\qquad do $1/2 \text{ step } v_z \rightarrow 1/2 \text{ step } v_y \rightarrow 1 \text{ step } v_x \rightarrow 1/2 \text{ step } v_y \rightarrow 1/2 \text{ step } v_z$
	
	\smallskip
	\quad Full position step:
	
	\qquad do $1 \text{ step } x \rightarrow 1 \text{ step } y \rightarrow  1\text{ step } z$
	
	\smallskip
	\quad Half velocity step using $\mathbf{E}^{t+1/2}$, $\mathbf{B}^{t+1/2}$:
	
	\qquad do $1/2 \text{ step } v_z \rightarrow 1/2 \text{ step } v_y \rightarrow 1 \text{ step } v_x \rightarrow 1/2 \text{ step } v_y \rightarrow 1/2 \text{ step } v_z$
	
	\bigskip
	
	Calculate $\mathbf{j}^{t+1}$ from the $f_i^{t+1}$
	
	Interpolate $\mathbf{j}^{t+1}$ to faces of Yee-cube
	
	\textbf{Full Maxwell update}
	
	\quad Advance $\mathbf{E}^{t+1/2}$ to $\mathbf{E}^{t+3/2}$ using $\mathbf{j}^{t+1}$
	
	\quad Advance $\mathbf{B}^{t+1}$ to $\mathbf{B}^{t+2}$, interpolate to $t+3/2$
}

\end{algorithm}

\subsection{Active Flux in One Dimension}\label{sec:AF}
\begin{figure}
\centering
\begin{subfigure}{0.33\textwidth}
	\includegraphics[width=\textwidth]{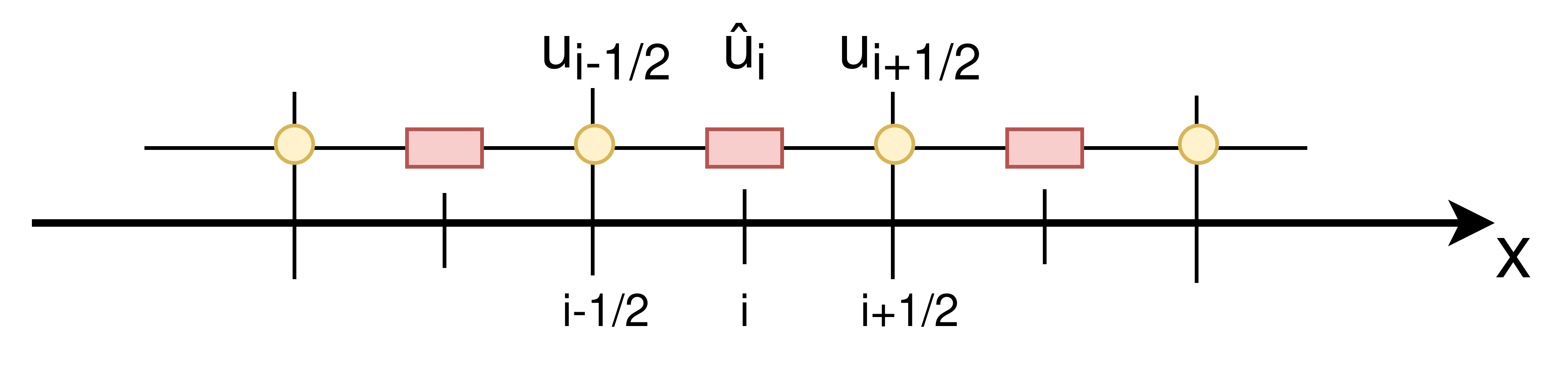}	
\end{subfigure}
\begin{subfigure}{0.3\textwidth}
	\includegraphics[width=\textwidth]{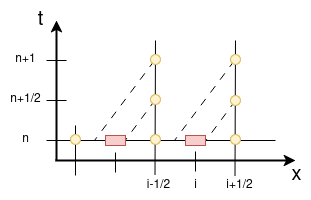}
\end{subfigure}
\begin{subfigure}{0.3\textwidth}
	\includegraphics[width=\textwidth]{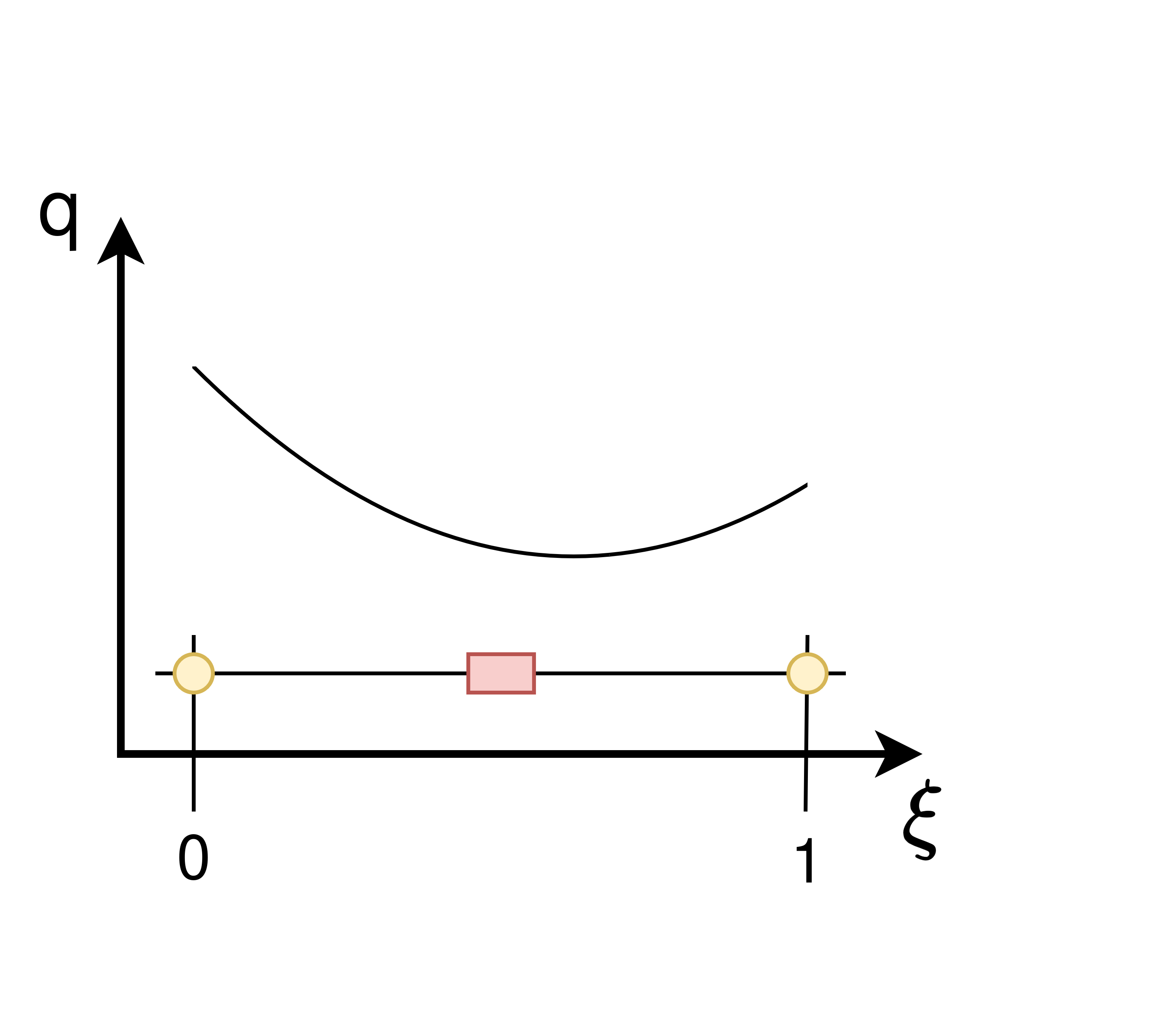}
\end{subfigure}
\caption{Left: Degrees of freedom in the 1D AF method. Center: Semi-Lagrangian update of the interface DoF by backtracing the characteristic. Right: Reconstruction of the solution to be evaluated at the footpoint of the characteristics.}
\label{fig:AF_basics}
\end{figure}

As described above, if we use splitting, then each equation during a respective splitting step corresponds to a one-dimensional advection equation with constant advection velocities. Therefore,
we consider the hyperbolic linear one-dimensional advection equation for the
conserved quantity $u(x,t) \in \mathbb{R}$ with the constant advection velocity $a \in
\mathbb{R}$:
\begin{align}
\partial_t u(x,t) + a \partial_x u(x,t) = 0.
\label{eqn:LinearAdvection_1D}
\end{align}
The solution is discretized into two types of degrees of freedom. The cell average of the ith cell $\hat{u}_i$ is stored at cell center, like it is custom for Finite Volume methods, but additionaly, a pointlike value of $u$ on the cell boundary $u_{i+1/2}$ is considered (see Fig. \ref{fig:AF_basics}, left panel):

\begin{align}
u_{i+1/2}^n &\approx u(x_{i+1/2}, t_n) \\
\la{u}_{i}^n &\approx \frac{1}{\Delta x} \int_{x_{i-1/2}}^{x_{i+1/2}} u(x,t^n) dx \; ,
\end{align} 
where, for simplicity, we assume a uniform spacing with $\Delta x = x_{i+1/2} - x_{i-1/2}$. 

Due to the cell boundary points $u_{i+1/2}$ being shared among neighboring cells, the number of cell internal DoF for the one-dimensional Active Flux method is two. The conservative update of the
cell average is performed as in classical Finite Volume Methods:
\begin{align}
\la{u}_i^{n+1} =  \la{u}_i^n + \frac{\Delta t}{\Delta x} [F_{i-1/2} - F_{i+1/2}].
\end{align} 
Here $F_{i+1/2}$ denotes the numerical fluxes through the interface, which are approximated with Simpson's rule in time to

\begin{align}
F_{i+1/2} &= \frac{a}{6} (u_{i+1/2}^n + 4u_{i+1/2}^{n+1/2} + u_{i+1/2}^{n+1}) \\ 
&\approx \frac{a}{\Delta t} \int_{t^n}^{t^{n+1}} u(x_{i+1/2},t) dt.
\label{eqn:FluxIntegration_1D}
\end{align}
To obtain each interface flux, the point DoF located at the interface needs to be evolved to half- and to full-time. This update does not need to be conservative. For the Vlasov equation, being a transport equation, it is straightforward to use semi-Lagrangian tracing of the characteristics (see Fig. \ref{fig:AF_basics}, center and right).

For this purpose, the following piecewise quadratic polynomial can be constructed from the interface and cell average values at time $t=t^n$:
\begin{align}
q_i(\xi) = u_{i-1/2} (\xi-1)(2\xi-1) + (6\la{u}_i - u_{i-1/2} - u_{i+1/2}) \xi(1-\xi) + u_{i+1/2} \xi (2\xi-1)
\label{eqn:Reconstruction_1D}
\end{align}     
with $\xi = (x-x_{i-1/2})/\Delta x$ and $0 \leq \xi \leq 1$. Due to the interface values being shared between neighbors, this reconstruction is continuous.

The reconstruction is then evaluated at the characteristic origins, which may be located in a neighboring cell:
\begin{align}
u_{i+1/2}^{n+1} = \begin{cases}
	q_{i}(\xi_R -\frac{a \Delta t}{\Delta x}) & a \geq 0 \\
	q_{i+1}(\xi_L -\frac{a \Delta t}{\Delta x}) & a < 0
\end{cases}
\label{eqn:TracingAdvection_1D}
\end{align}
Here $\xi_R=1$ corresponds to the right interface value and $\xi_L=0$ to the
left. The values at half-time $u_{i+1/2}^{n+1/2}$ can be obtained analogously.
The Courant number $\nu = (a\Delta t)/\Delta x$ is hereby restricted by the
usual CFL condition for stability $0 \leq |\nu| \leq 1$. For the considered case
of a constant velocity advection we can carry out the mentioned steps easily by
explicit calculation which yields the following update equations for interface
and cell average values, respectively:
\begin{align}
u_{i+1/2}^{n+1} = \begin{cases}
	\nu(3\nu-2) u_{i-1/2}^n + 6\nu (1-\nu) \la{u}_i^n + (1-\nu)(1-3\nu) u_{i+1/2}^n & a \geq 0 \\
	(1-\nu)(1-3\nu)u_{i-1/2}^n + 6\nu (1-\nu) \la{u}_i^n + \nu(3\nu-2)u_{i+1/2}^n & a < 0
\end{cases}
\label{eqn:UpdateFormula_1D_Interface}
\end{align}

\begin{align}
\la{u}_i^{n+1} = \begin{cases}
	\nu^2 (\nu-1) u_{i-3/2}^n + \nu^2(3-2\nu) \la{u}_{i-1}^n + \nu(1-\nu) u_{i-1/2}^n + (1-\nu)^2 (1+2\nu) \la{u}_{i}^n - \nu(1-\nu)^2 u_{i+1/2}^n & a \geq 0 \\
	\nu^2 (\nu-1) u_{i+3/2}^n + \nu^2(3-2\nu) \la{u}_{i+1}^n + \nu(1-\nu) u_{i+1/2}^n + (1-\nu)^2 (1+2\nu) \la{u}_{i}^n - \nu(1-\nu)^2 u_{i-1/2}^n & a < 0
\end{cases}
\label{eqn:UpdateFormula_1D_Average}
\end{align}
It should be stressed that these update formulas are both easy to implement and computationally economical with respect to computing time and memory usage, which is of utmost importance when dealing with high-dimensional problems. 

\subsection{The Split-Step Active Flux Grid}

\begin{figure}[ht]
\centerline{\includegraphics[scale=0.65]{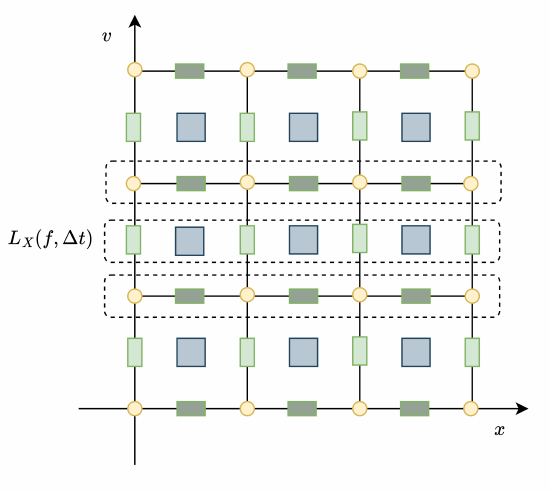}
	\hspace*{1cm}
	\includegraphics[scale=0.65]{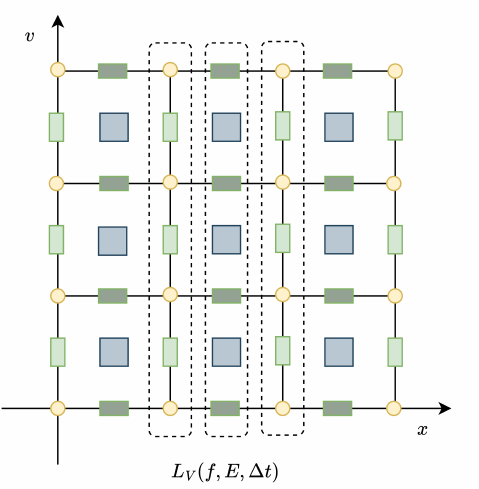}
}
\caption{Splitting approach for a 2D AF cell in phase space. The evolution operators for the update in $x-$direction (left) and  $v-$direction (right) are applied on (one-dimensional) fibers of the grid along each direction.
	For the 2D grid, this means the following: the corner DoF are point values and the center DoF are cell averages, but edge DoF are treated as line averages in direction of the respective edge. This is the result of a given edge value acting as the central average during the update in the direction parallel to its edge, but being treated as an interface point value during the update in normal direction. Figure as in \cite{Hensel2024}.}
\label{fig:2nd_order_naive}
\end{figure}

The 1D AF method is sequentially applied fiber-wise to the higher-dimensional grid like in \cite{Hensel2024}, which is exemplarily sketched in Fig.~\ref{fig:2nd_order_naive} for the 2D case. As a result, the higher-dimensional grid needs to be inhomogeneously composed of different types of averages to accommodate the requirement of 1D AF to alternate point values and line integrals \emph{with respect to the update direction}.

\begin{figure}[h]
\centering
\includegraphics[width=0.5\textwidth]{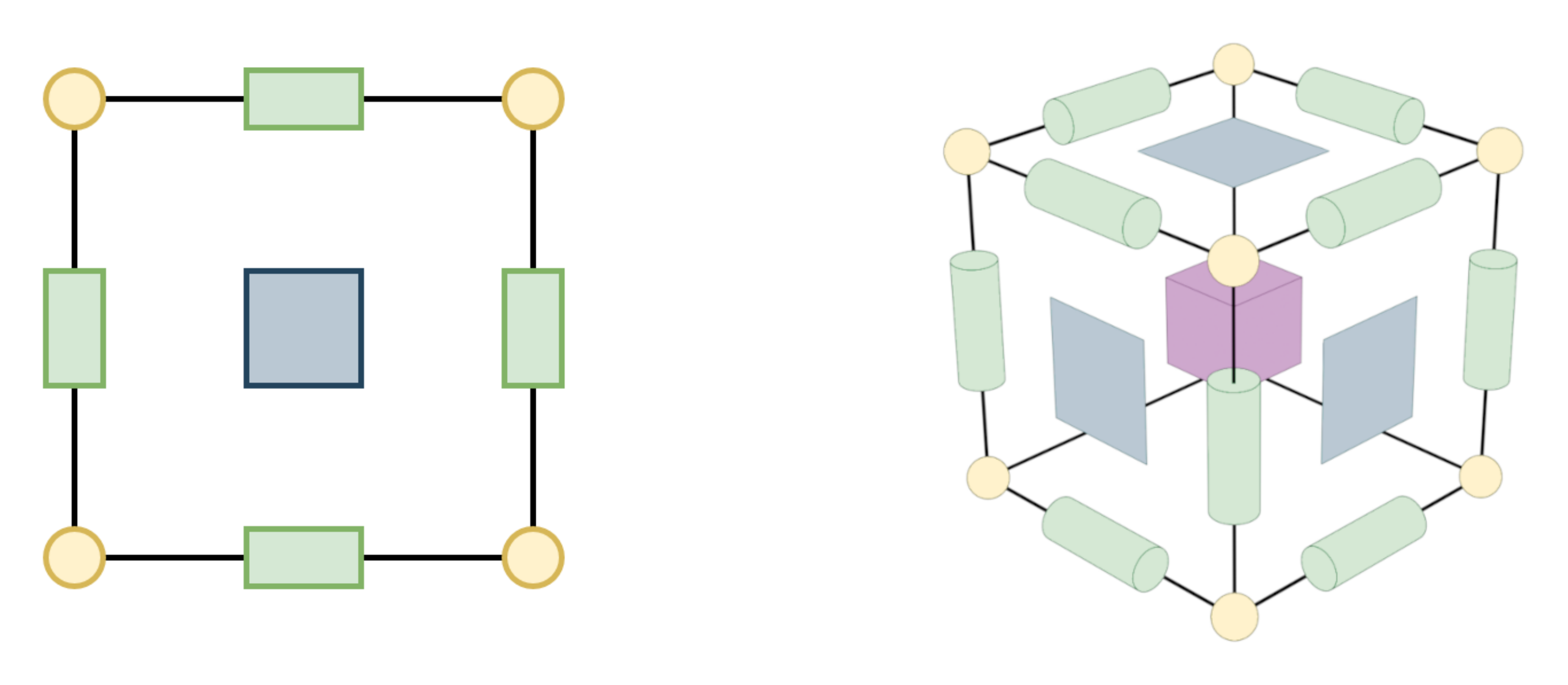}
\caption{Cells of the split-step AF grid. Left: 2D. Corner values remain point values and center values remain cell averages, but edge values are line averages depending on their orientation. Right: 3D. The grid consists of point values on the cell corners (yellow), edge averages (green), face averages (blue), and the cell average (purple).}
\label{fig:grid}
\end{figure}

This is realized in $d$ dimensions by including cell averages ($d$-forms) as well as face averages ($k$-forms) on each $k$-face of the cell, $k=0,\ldots, d-1$. In 2D, this results in the grid shown in the left panel of Fig.~\ref{fig:grid}, with point values on the corners, cell-integrals at the center, and line-averages on each cell interface. 
The realization in 3D is shown in the right panel of Fig.~\ref{fig:grid}, where points are marked as balls, line-integrals on edges as cylinders, 2D integrals over the facets (2-faces) as squares and, the cell integral as a cube. In general, for each 1D line intersection the $d$-dimensional cube, we then have a $k$-form on the interface and a $k+1$-form (adding the line integral along the considered direction).

The update of the cell average is then only correct to second order, as the following approximation is made to the numerical fluxes by using values of advection velocity and solution which are integrated over the respective cell interface to evaluate the flux function, instead of evaluating it from point values and thereafter integrating over the interface:
\begin{align}\label{eq:doubleintegral}
&\frac{1}{\Delta t} \prod_{i\neq j} \frac{1}{\Delta w_i}\int_{t^n}^{t^{n+1}} \int_{w_{i,k-1/2}}^{w_{i,k+1/2}} a(\mathbf{w},t) \cdot f(\mathbf{w},t)\,  d^5w_{i\neq j}\, dt \\
&= \frac{1}{\Delta t} \prod_{i\neq j} \frac{1}{\Delta w_i}\int_{t^n}^{t^{n+1}}\left( \int_{w_{i,k-1/2}}^{w_{i,k+1/2}} a(\mathbf{w},t)\,  d^5w_{i\neq j} \cdot \int_{w_{i,k-1/2}}^{w_{i,k+1/2}} f(\mathbf{w},t)\, d^5w_{i\neq j} \right) dt + \mathcal{O}(\max_{i\neq j}{\Delta w}^2) \; ,
\end{align}
where $\mathbf{w} = (\mathbf{x},\mathbf{v}) \in \mathbb{R}^6$ denotes the entire phase space, $a(\mathbf{w},t)$ the advection speed, and $j$ is the updated direction.
Specifically, this order reduction arises during the velocity updates of the cell averages, where the advection velocity $a=\mathbf{E}(\mathbf{x})+\mathbf{v}\times\mathbf{B}(\mathbf{x})$ indeed depends on the location in configuration space.

A higher-order approximation, which however is more costly, can be found in \cite{Hensel2024} for the 1D1V Vlasov--Poisson case.

\subsection{Positive and Flux Conservative Method}
The Positive and Flux Conservative Method (PFC), introduced in \cite{Filbet2001}, is a conservative method based on the semi-Lagrangian principle using slope limiters to ensure positivity of $f$. It is an established third-order accurate method for the Vlasov problem. Here, only the cell average is stored and its neighbors are used to reconstruct the solution at the characteristic origins. To do this correctly to third order, the stencil thus cannot be compact while including sufficient DoF. The PFC method is technically not restricted by a CFL condition. In practice however, we constrain its stencil to the two neighboring cells in both directions. This way, the number of DoF to be exchanged is the same as for Active Flux at half the resolution i.e. same number of DoF to be evolved. Consequentially, the timestep is restricted by a CFL $\le1$. This means that on the other hand, at the same number of DoF, the time step used for Active Flux can be twice as large as the one used for PFC.
PFC is known to be dissipative, more so than AF \cite{Hensel2024}.

\subsection{Finite-Difference Time-Domain Method}
\begin{figure}[h]
\centering
\begin{subfigure}{0.3\textwidth}
	\includegraphics[width=\textwidth]{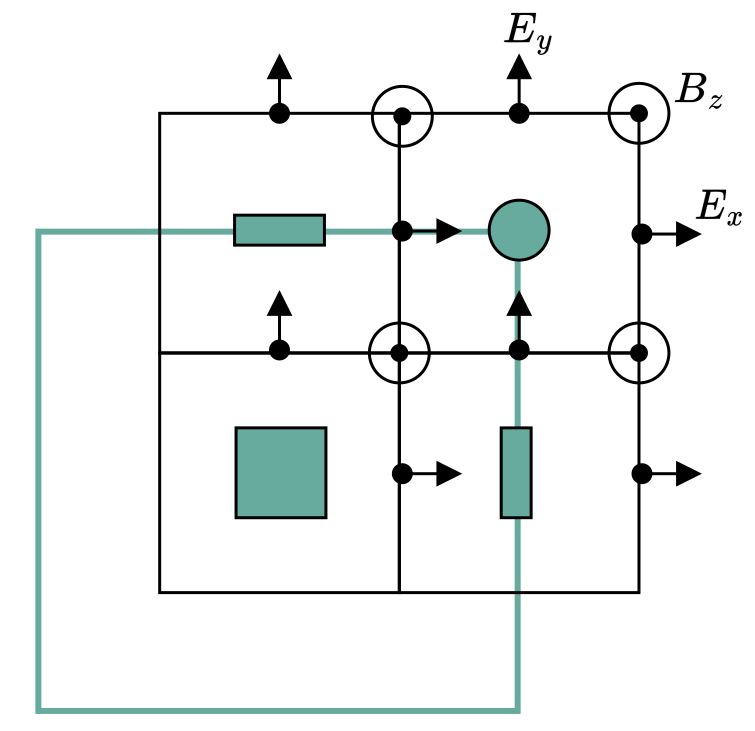}
	\caption{AF}
\end{subfigure}
\hspace{0.1\textwidth}%
\begin{subfigure}{0.15\textwidth}
	\includegraphics[width=\textwidth]{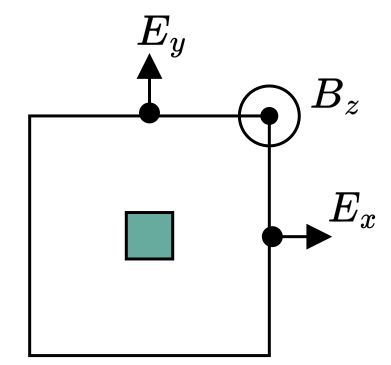}
	\caption{PFC}
\end{subfigure}
\caption{A slice at the location of the respective DoF in configuration space of AF and PFC, respectively, as well as the chosen Yee-grid. For PFC, the Yee-cube coincides with the configuration space cell, while in the AF case, each DoF hosts its individual Yee-cube.}
\label{fig:Yee}
\end{figure}

Maxwell's equations are evolved using the Finite-Difference Time-Domain (FDTD) method. Here, the degrees of freedom are arranged on the Yee-cube with the components of $\mathbf{j}$ and $\mathbf{E}$ being located on its faces, the components of $\mathbf{B}$ on its edges, and each being the corresponding average \cite{Yee1966}. This conforms with the geometric properties of Maxwell's equations, and the Yee method can be interpreted as a second-order case of exterior-calculus-based methods (see e.g \cite{Teixeira2013}). As a result, $\nabla\cdot \mathbf{B}=0$ is ensured to be conserved. In time, the FDTD method employs a Leapfrog integrator.
For the PFC scheme, $\mathbf{j}$ is calculated at the cell center, where the cell average of $f_s$ is stored and then linearly interpolated to the cell face.
The additional degrees of freedom in AF are exploited by solving the fields for each of the consequently available $\mathbf{j}$. In other words, a Yee-cube is placed around each AF DoF, regardless of their nature, and - analogously to the PFC case - $\mathbf{j}$ is calculated at the center of the Yee cube and interpolated to its face. In this sense, the Yee-grid constructed for AF is equivalent to the one considered for PFC at twice the resolution. This is illustrated in \autoref{fig:Yee}, where the Yee-grid is related to the grids of AF and PFC, respectively. 

For the computation of $\mathbf{j}$, the integral over the velocity space is performed by summing over all available DoF. This approximates the desired Yee-face average of $\mathbf{j}$ to second order.
Poisson's equation is discretized using the standard second-order finite difference stencil. A Conjugate Gradient (CG) solver is used to solve the resulting linear system. 

\subsection{Boris correction}
The error in Gauss' law arising from the discretization of the system can cause unphysical electric fields. Particularly at low resolutions, this can ultimately cause the simulation to become instable. For methods performing a FV update, the fluxes can be used to calculate the current density such that it exactly fulfills the discrete charge continuity equation as to obtain a consistent electric field \cite{crouseilles2014}. Despite using a FV update for the DoF at cell center, this approach is not straightforwardly possible for the AF scheme, as we solve Maxwell's equations not once per cell, but on a Yee cube centered around each DoF, including the non-conservatively updated interface values. In this work, we therefore apply the classic Boris correction (see e.g. \cite{Birdsall}), correcting the electric field $\mathbf{E}$ retroactively. The correction term added to $\mathbf{E}$ at each timestep is found by solving Poisson's equation (see Alg. \ref{alg:timestepping_boriscorr}), which is done as described in the previous section. The Boris correction will be added to the scheme when signified and with a timestepping as in Alg. \ref{alg:timestepping_boriscorr}.

\SetKwBlock{Timestep}{Time step}{end}
\begin{algorithm}
\caption{Time stepping including a Boris correction}\label{alg:timestepping_boriscorr}
\textbf{Initialize }

\smallskip
Half Maxwell update:

\quad Advance $\mathbf{E}^{t=0}$ to $\mathbf{E}^{t+1/2}$ using $\mathbf{j}^{t=0}$

\quad Advance $\mathbf{B}^{t=0}$ to $\mathbf{B}^{t+1}$, interpolate to $t+1/2$

\Timestep{
	Calculate $\mathbf{\rho}^{t+1}$ from the $f_i^{t}$
	
	\medskip
	Interpolate fields to center of Yee-cube
	
	\textbf{Full Vlasov update}
	
	\quad Half velocity step using $\mathbf{E}^{t+1/2}$, $\mathbf{B}^{t+1/2}$:
	
	\qquad do $1/2 \text{ step } v_z \rightarrow 1/2 \text{ step } v_y \rightarrow 1 \text{ step } v_x \rightarrow 1/2 \text{ step } v_y \rightarrow 1/2 \text{ step } v_z$
	
	\smallskip
	\quad Full position step:
	
	\qquad do $1 \text{ step } x \rightarrow 1 \text{ step } y \rightarrow  1\text{ step } z$
	
	\smallskip
	\quad Half velocity step using $\mathbf{E}^{t+1/2}$, $\mathbf{B}^{t+1/2}$:
	
	\qquad do $1/2 \text{ step } v_z \rightarrow 1/2 \text{ step } v_y \rightarrow 1 \text{ step } v_x \rightarrow 1/2 \text{ step } v_y \rightarrow 1/2 \text{ step } v_z$
	
	\bigskip
	Boris correction of  $\mathbf{E}^{t+1/2}$:
	
	\quad Calculate ${\rho}^{t+1}$ from the $f_i^{t+1}$, \quad ${\rho}^{t+1/2} = ({\rho}^{t+1}+{\rho}^{t})/2$

	\quad Solve $\Delta \delta\Phi = \nabla \cdot \mathbf{E}^{t+1/2} -\rho^{t+1/2}/\epsilon_0 $
	
	\quad Obtain $\mathbf{E}_{corr}^{t+1/2}=\mathbf{E}^{t+1/2}-\nabla\delta\Phi$
	
	\bigskip
	
	Calculate $\mathbf{j}^{t+1}$ from the $f_i^{t+1}$
	
	Interpolate $\mathbf{j}^{t+1}$ to faces of Yee-cube
	
	\textbf{Full Maxwell update}
	
	\quad Advance $\mathbf{E_{corr}}^{t+1/2}$ to $\mathbf{E}^{t+3/2}$ using $\mathbf{j}^{t+1}$
	
	\quad Advance $\mathbf{B}^{t+1}$ to $\mathbf{B}^{t+2}$, interpolate to $t+3/2$
	
}

\end{algorithm}

\section{Numerical Experiments}
The proposed scheme was implemented in the  \textit{muphyII}-code \cite{Allmannrahn2024, lautenbach_2024_10547265}, which is a multiphysics framework for collisionless plasma simulations, targeting space plasmas in particular. It allows for dynamic and adaptive interface coupling of different plasma models, but can also be utilized for simulations of a specific model, like it was done for the present numerical tests of the split-step AF Vlasov solver. The PFC Vlasov solver of the same framework was used as a benchmark. The code employs MPI decomposition to the physical domain. 

\subsection{Normalization}
For the electrostatic setups in \autoref{sec:BW} and \autoref{sec:BNS}, all quantities are scaled with respect to characteristic electron parameters, as they focus on electron dynamics in the presence of a static ion background.  
Lengths are normalized by the electron skin depth $d_{e,0}$, which is defined based on the reference density $n_0$. 
Velocities are expressed in units of the vacuum speed of light $c_{0}$, while time is scaled by the inverse of the electron plasma frequency $\omega_{p,0}^{-1}$. 
Masses are normalized by the electron mass $m_e$.  The vacuum permeability is set to $\mu_0 = 1$ and the vacuum permittivity to $\epsilon_0 = 1$. This implies $c_0=1/\sqrt{\epsilon_0\mu_0}=1$.
Furthermore, the Boltzmann constant is taken as $k_B = 1$. Note that this is equal to normalizing lengths by the electron Debye length and velocities by the electron thermal velocity, if electron temperature and mass are set to one, which is the case for the Landau damping setups \autoref{sec:LD}. 

On the other hand, the Orszag-Tang vortex setup (\autoref{sec:OT}) uses ion-based normalization.
Here, the characteristic length is the ion inertial length $d_{i,0}$, again defined from the reference density $n_0$. 
Velocities are normalized by the ion Alfvén speed $v_{A,0}$, which depends on the reference magnetic field $B_0$. 
Time is scaled by the inverse of the ion cyclotron frequency $\Omega_{i,0}^{-1}$, and masses are expressed in units of the ion mass $m_i$. 
The vacuum permeability is set to $\mu_0 = 1$ and the Boltzmann constant is again taken as $k_B = 1$.

\subsection{Spherical Blast Wave}\label{sec:BW}
\begin{figure}
\centering
\includegraphics[width = 1.\textwidth]{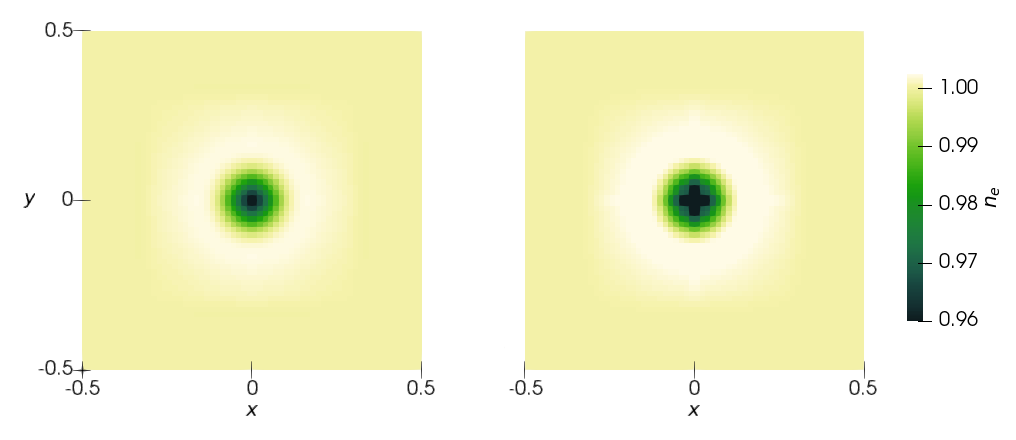}
\caption{Blast wave test at $t=0.05$. Left: AF with $64^2\times32^2$ cells, right: PFC with $64^2\times32^2$.}
\label{fig:blast}
\end{figure}

Inspired by the Taylor--von Neumann--Sedov solution of Euler's equations (used for testing purposes e.g. in \cite{Schaal2015}), we simulated a radially expanding blast wave caused by a pointlike source of energy in an electron plasma of uniform density in 2D2V. The electron distribution was initialized as
\begin{align}
f_{e, 0}(\mathbf{x}, \mathbf{v})&=
\begin{cases}
	\frac{1}{2 \pi T_{s} } \exp\left({\frac{-\mathbf{v}^2}{2T_{s}}}\right)& \text{a square region of size $2/64^2$ around the origin \footnotemark}\\
	\frac{1}{2 \pi T_{bg}} \exp\left({\frac{-\mathbf{v}^2}{2T_{bg}}}\right)	& \text{anywhere else,}
\end{cases}	
\end{align}
with a source electron temperature $T_{s}=10$ and a background electron temperature $T_{bg}=1$.\footnotetext{In the case of AF, the cell-averages of these innermost cells as well as all interface values in between were considered.}
The Ions were taken to be a static and uniform neutralizing background, and the fields were initialized to zero. The computational domain extends to $\Omega=[-0.5,0.5]^2 \times[-6,6]^2$. As shown in Fig. \ref{fig:blast}, the breach of the spherical symmetry is much more pronounced in the PFC simulation. The comparison is made on basis of the number of cells (like in \cite{Schaal2015}), and the increased memory requirement per cell of AF should be acknowledged.
\subsection{Linear Landau Damping}\label{sec:LD}
Landau damping is a prominent damping mechanism of waves in collisionless plasmas and, as an analytic damping rate can be obtained through linear theory of the Vlasov equation, a standard test for kinetic solvers. 

To induce linear Landau damping in 3D3V, we initialized the electron distribution function as a Maxwellian with a low amplitude ($\alpha=0.01$) perturbation in each spatial direction and the fields accordingly:
\begin{align}
f_{e, 0}(\mathbf{x}, \mathbf{v})&=\frac{1}{(2 \pi)^{3 / 2}} \exp\left({\frac{-\mathbf{v}^2}{2}}\right)\left(1+\alpha \sum_{j=1}^3 \cos \left(k_j x_j\right)\right)\\
E_i &= -2\alpha\sin(k_ix_i), \text{ and } B_i = 0,
\end{align}
where $k_i=0.5$ and the phase space extends to $\Omega=[-2\pi,2\pi]^3 \times[-6,6]^3$. It was set $c_0=30$. The timestep was chosen according to $\nu=\text{max}(\mid v_e\mid)\Delta t / \Delta x = 0.8$. The computation of the electromagnetic fields was subcycled with 15 subcycles to relax the CFL-condition of the Maxwell solver. The ions were again static and uniform in space as a neutralizing background.

\begin{figure}
\centering
\includegraphics[width = 0.9\textwidth]{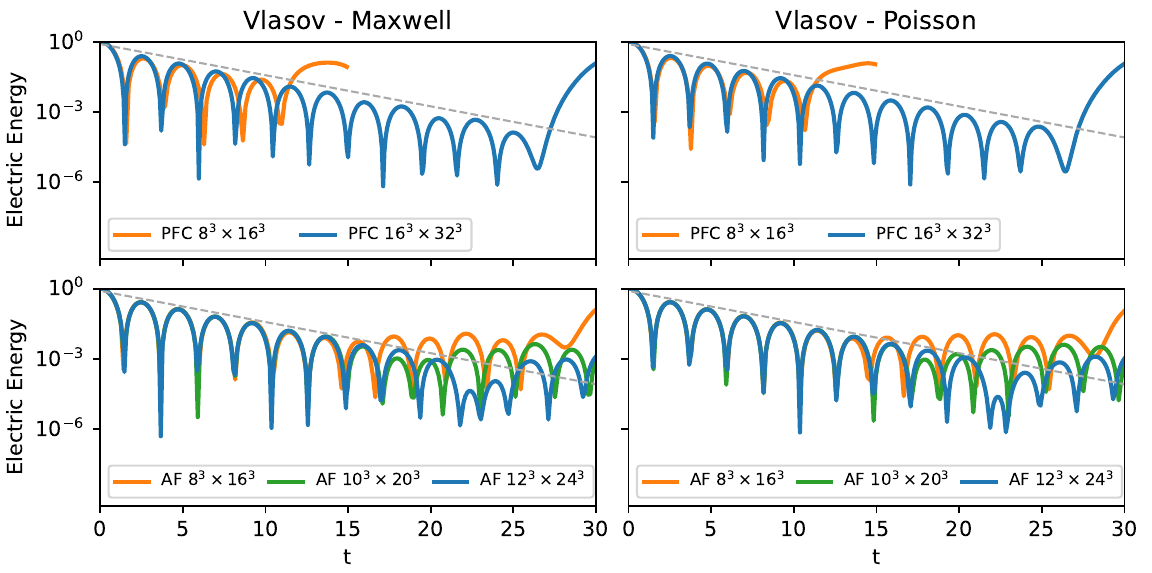}
\caption{Landau damping simulation using the Vlasov--Maxwell schemes (left column), and the Vlasov--Poisson schemes (right column). The upper row are the results for the PFC solver at two (cell-based) resolutions. The lower row is AF, where the lowest resolution has the same number of cells as the low resolution PFC run, but the same number of DoF as the high resolution PFC run. Additional intermediate resolutions are shown. The gray dashed line is the analytically expected energy damping rate $\gamma=0.3066$.}
\label{fig:LD}
\end{figure}

In \autoref{fig:LD}, we compare the simulation results for the AF-based Vlasov--Poisson and Vlasov--Maxwell schemes to the PFC-based counterparts. The AF simulations perfectly capture the analytically expected damping rate, PFC shows slightly increased damping in the time interval $[0,15]$. At the same number of grid cells, the resolution of the phase-space filamentation by AF is superior to that of PFC, which manifests in the slightly postponed recurrence time. This is to be expected due to the additional sampling at the cell interfaces in AF. The final stages of the damping phase just before the onset of recurrence are however not as accurately reproduced, even when moving towards intermediate grid cell numbers.

\begin{figure}
\centering
\includegraphics[width = 0.9\textwidth]{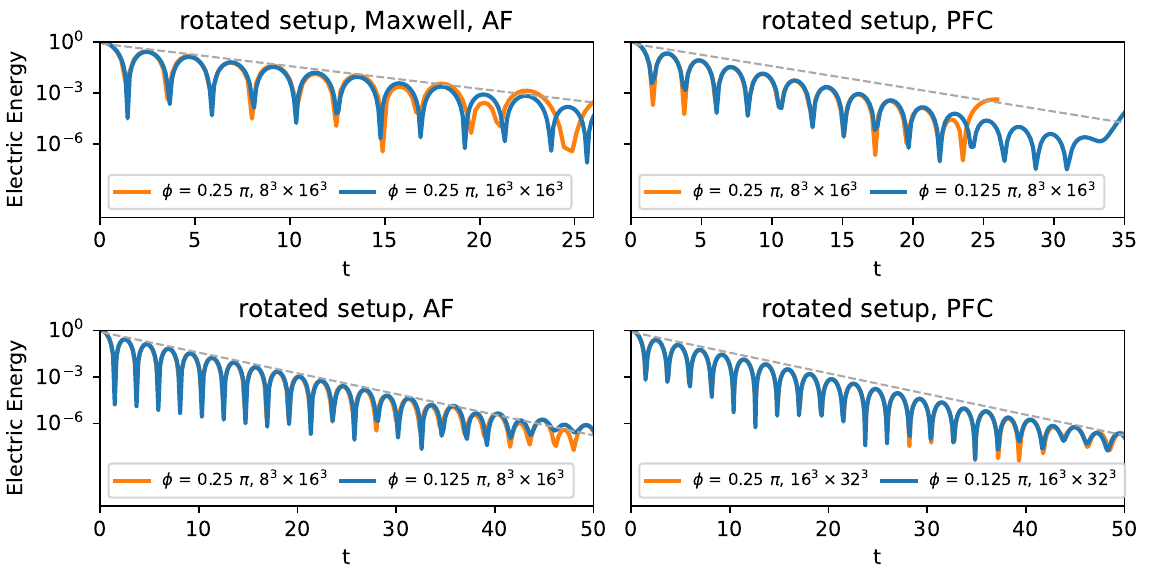}
\caption{Rotated Landau damping setup for the Vlsaov--Maxwell AF scheme (upper left panel), the Vlasov--Poisson AF scheme (lower left panel), and the Vlasov--Poisson PFC scheme at the same number of cells as its AF counterpart (upper right panel), and at the same number of DoF (lower right panel). The gray dashed line is again the analytic solution.}
\label{fig:LD2}
\end{figure}

In the Vlasov--Maxwell simulation of the problem (which is actually sufficiently described in the Poisson limit), we notice the very good reproduction of the electrostatic field by the Maxwell solver, undisturbed by spurious electromagnetic modes, despite the low resolutions. There is however a slight irregularity of phases caused by the spatial discretization of the perturbation towards the end of the simulation time, which is not visible in the Vlasov--Poisson runs. This becomes more pronounced when the perturbation is no longer mesh-aligned and isotropic (\autoref{fig:LD2}, upper left panel). Here, we initialized a single perturbation rotated by a small angle $\phi=\{\pi/4, \pi/8\}$ about two axes. The physical domain was scaled such that it remains periodic. The Yee scheme suffers from numerical anisotropy due to the field components being considered at different sides of the Yee cube. For this reason, in applications such as material sciences, modified stencils reducing the anisotropy error are often used (see e.g. \cite{Sescu2015} for a review and \cite{Sekido2024} for a recent contribution). Even though this problem is alleviated by higher spatial resolutions, we make the comparison of the Vlasov solvers using Poisson's equation.
We notice that the frequency resolution of AF is closer to PFC at the same number of DoF than grid cells. Moreover, the dissipation of the wave amplitude in the PFC simulations becomes significant in the ongoing simulation. Due to the rotation of the initial perturbation, the recurrence now occurs at later times.


\subsection{Electron Bernstein Waves}\label{sec:BNS}
The dispersion relation of electromagnetic waves propagating perpendicular to a constant uniform magnetic field in warm plasma is, as can again be found through linearization:
\begin{align}\operatorname{det}\left(\begin{array}{cc}1-\mathrm{M} \frac{\mathrm{e}^{-\lambda}}{\lambda \omega} \sum_{n=-\infty}^{\infty} A_n & \mathrm{iM} \frac{\mathrm{e}^{-\lambda}}{\omega} \sum_{n=-\infty}^{\infty} B_n \\ -\mathrm{iM} \frac{\mathrm{e}^{-\lambda}}{\omega} \sum_{n=-\infty}^{\infty} B_n & 1-\frac{\lambda}{\omega^2} \frac{c^2}{\mathrm{M}^2 v_{\mathrm{T}}^2}-\mathrm{M} \frac{\mathrm{e}^{-\lambda}}{\lambda \omega} \sum_{n=-\infty}^{\infty} C_n\end{array}\right)=0,
\end{align}
where $A_n=n^2 I_n(\lambda) /(\omega \mathrm{M}-n)$, $B_n=\left(n\left(I_n^{\prime}(\lambda)-I_n(\lambda)\right)\right) /(\omega \mathrm{M}-n)$, and $C_n=\left(n^2 I_n(\lambda)+2 \lambda^2 I_n(\lambda)-\allowbreak 2 \lambda^2 I_n^{\prime}(\lambda)\right) /(\omega \mathrm{M}-n)$. $I_n$ denotes the modified Bessel function of the first kind, $v_{\mathrm{T}}$ the thermal velocity, $\lambda= k^2 T_e/(m_e \Omega^2)$, $M=1/\Omega$, and $T_e$ and $m_e$ the electron temperature and mass, respectively \cite{Li2019,Stix1992}.

As opposed to the case of cold plasmas, where the gyroradius tends to zero, the dispersion relation is now affected by wave-particle interaction at the fundamental but also the higher harmonic gyrofrequencies, with the contribution of the latter increasing as the gyroradius approaches the magnitude of the wavelength. 

Here, the electric field $\mathbf{E}$ lies in the $xy$-plane, corresponding to the extraordinary modes in cold plasmas. If $\mathbf{E}$ is parallel to $\mathbf{k}$, the wave is nearly electrostatic and is called Bernstein wave \cite{Bernstein1958}.

\begin{figure}[h]
\centering
\begin{subfigure}{0.45\textwidth}
	\includegraphics[width=\textwidth]{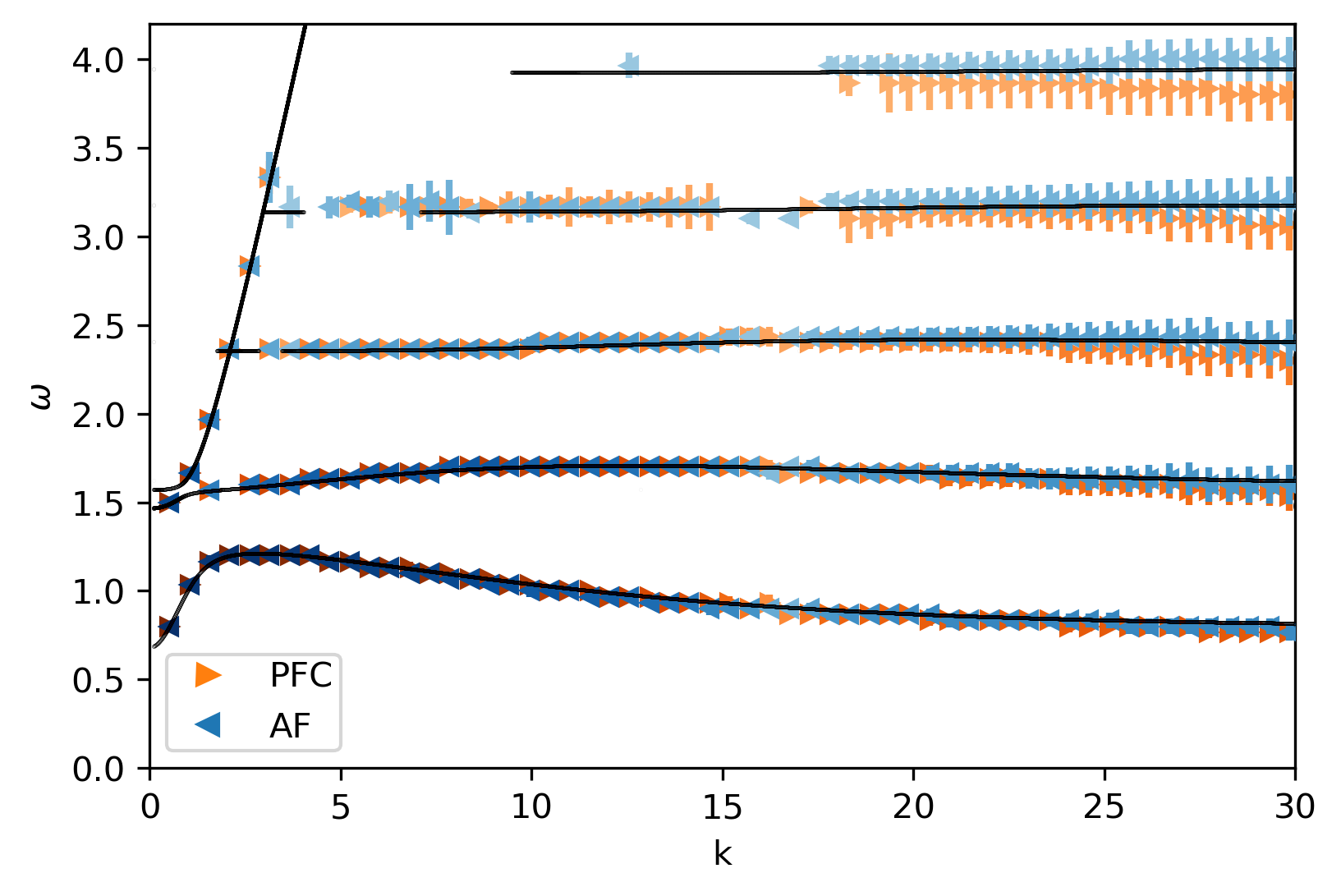}
\end{subfigure}
\begin{subfigure}{0.45\textwidth}
	\includegraphics[width=\textwidth]{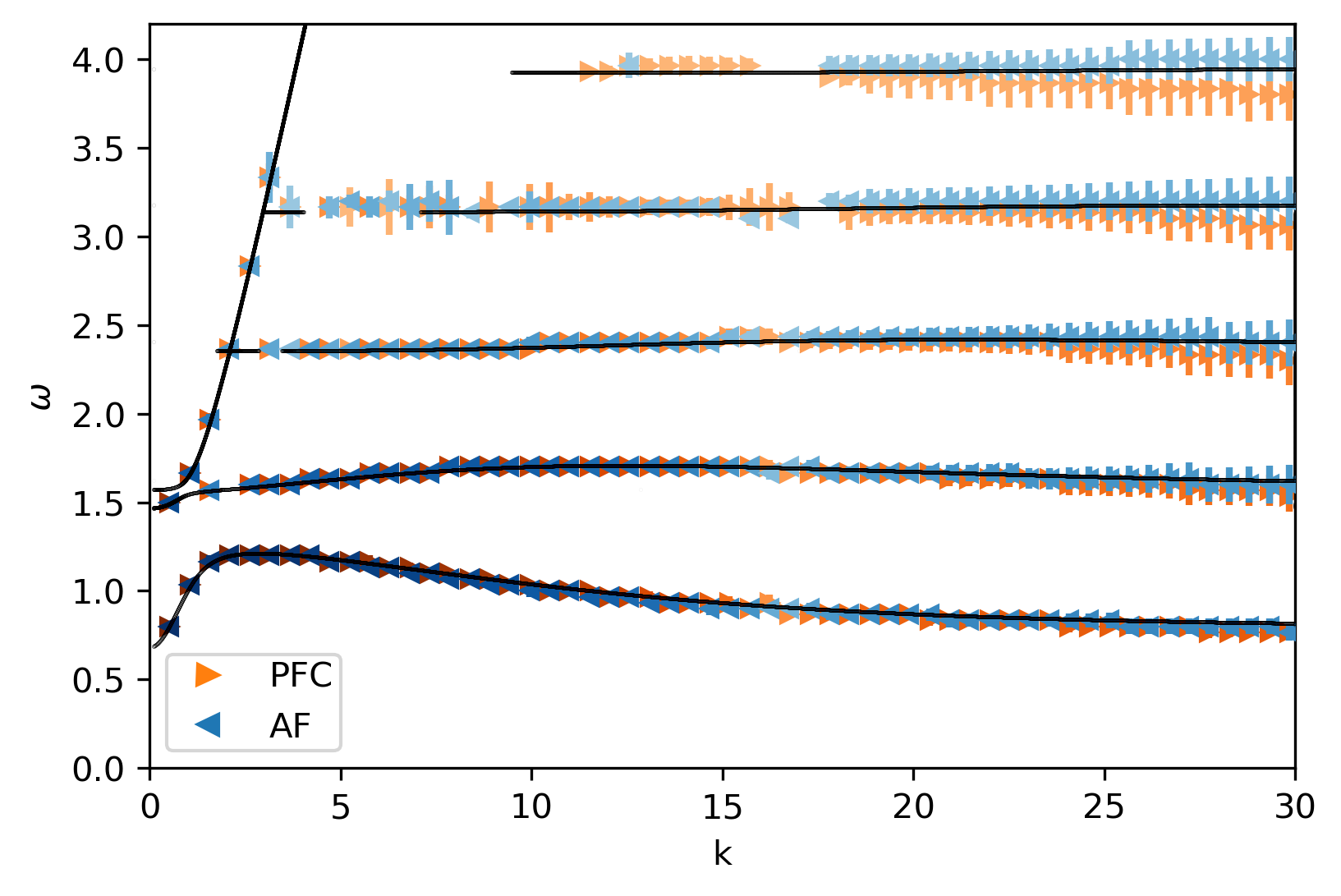}
\end{subfigure}
\caption{Dispersion relation of the electron Bernstein modes. Color depth indicates spectral amplitude. The black line is the analytical solution. Left: PFC with its limiter, right: PFC without it.}
\label{fig:bns}
\end{figure}

Despite the electrostatic nature of the problem, we again considered the Vlasov--Maxwell system with a static and uniform ion background, this time in one spatial and two velocity dimensions.
As in \cite{Li2019}, we initialized the electron distribution function and the fields as
\begin{align}
f\left(x, v_x, v_y\right)=\frac{1}{2 \pi T_e} \exp\left({-\frac{ v_x^2+v_y^2}{2 T_e}}\right)\left(1+a \sum_{n=1}^s n \cos \left(\frac{\pi}{6} n x\right)\right), & \\
E_x=\frac{6 a}{\pi} \sum_{n=1}^s \sin \left(\frac{\pi}{6} n x\right), \quad E_y=0, \quad \text { and } \quad B_z=0.785, &
\end{align}
with a phase space domain size of $\Omega=[0,12] \times[-0.4,0.4] \times[-0.4,0.4]$. $T_e$ was set to 0.0036. $s=60$ modes were initialized and the amplitude of the ground mode of the perturbation was $a=10^{-7}$. During the simulation time of $60\pi$, $E_x$ was documented after each time step to obtain the space-time spectrum of the wave. While the domain is periodic in space, this can not be assumed for the time, and therefore a Tukey window (width $\alpha=0.05\%$) was applied to the data along the time axis before performing the Fourier transform. To allow for a quantitative comparison between the schemes, the amplitude maxima along the $\omega$-axis were extracted, and the widths of the peaks were employed as error bars. In \autoref{fig:bns}, we compare the analytical dispersion relation to the space-time spectrum obtained from simulations using PFC at a resolution of $512\times64^2$ grid cells, and AF resolved by $256\times32^2$ grid cells. For AF, the additional point values were sampled in addition to the central average so that the spatial sampling rate was the same for both simulations. The timestep was $0.006$ for AF and $0.003$ for PFC. The latter was granted the doubled temporal sampling rate by recording the extra time steps. Still, AF outperformed PFC in the resolution of high wavenumbers and fast modes. Due to the long simulation time, the dissipation error is here dominant. This was also the case when the PFC limiter was turned off, i.e. reducing it to the conservative semi-Lagrangian method. The increased dissipation is therefore likely due to the larger stencil, meaning a larger area is averaged over to compute the fluxes in PFC.

\subsection{Orszag-Tang Vortex}\label{sec:OT}

The Orszag-Tang vortex \cite{Orszag1979} initializes decaying plasma turbulence and is a popular test case for kinetic plasma solver as it depends on the correct reproduction of several kinetic effects including Landau damping, magnetic reconnection, and the propagation of different wave modes to give rise to the turbulent energy dissipation.

Similarly as in \cite{Groselj2017}, we set up
\begin{align}
B_x&=-\delta_B \sin(2\pi y/L) B_0,& B_y&=\delta_B \sin(4\pi y/L) B_0,& B_z&=B_0 \nonumber \\
E_x&=-\delta_u B_0\sin(2\pi x/L),& E_y&=-\delta_u B_0\sin(2\pi y/L),& E_z&=0  \nonumber\\
u_{x,s}&=-\delta_u \sin(2\pi y/L),& u_{y,s}&=\delta_u \sin(2\pi x/L),& u_{z,i}&=0,\nonumber\\
&\text { and } u_{z,e}=-\frac{2\pi}{L}\delta_B\mu_0(2\cos (4\pi x/L)+\cos(2\pi y/L))\mkern-36mu \mkern-36mu \mkern-36mu \mkern-36mu \mkern-36mu\mkern-36mu&&& \end{align}
over a 2D physical domain of size $L=L_x=L_y=4\pi$. The perturbation magnitudes were $\delta_u=0.2$ and $\delta_B=0.2$. Further, it was $\beta_i=2 T_i / B_0^2=0.1$, $T_i=T_e=0.05$, $m_i/m_e=25$, and $c=18.174$.
The distribution functions $f_s$ were Maxwellian in $\mathbf{v}$ accordingly, extending to $\pm 7$ for the electrons and $\pm 1.5$ for the ions.

At very low resolutions, the violation of Gauss' law becomes significant \cite{Birdsall}, in particular where sharp gradients are present, and needs to be corrected to suppress spurious oscillations and stabilize the AF scheme. For the PFC scheme, where these gradients are not captured to begin with due to dissipation, this is not necessary.

\begin{figure}
\centering
\includegraphics[width = 0.9\textwidth]{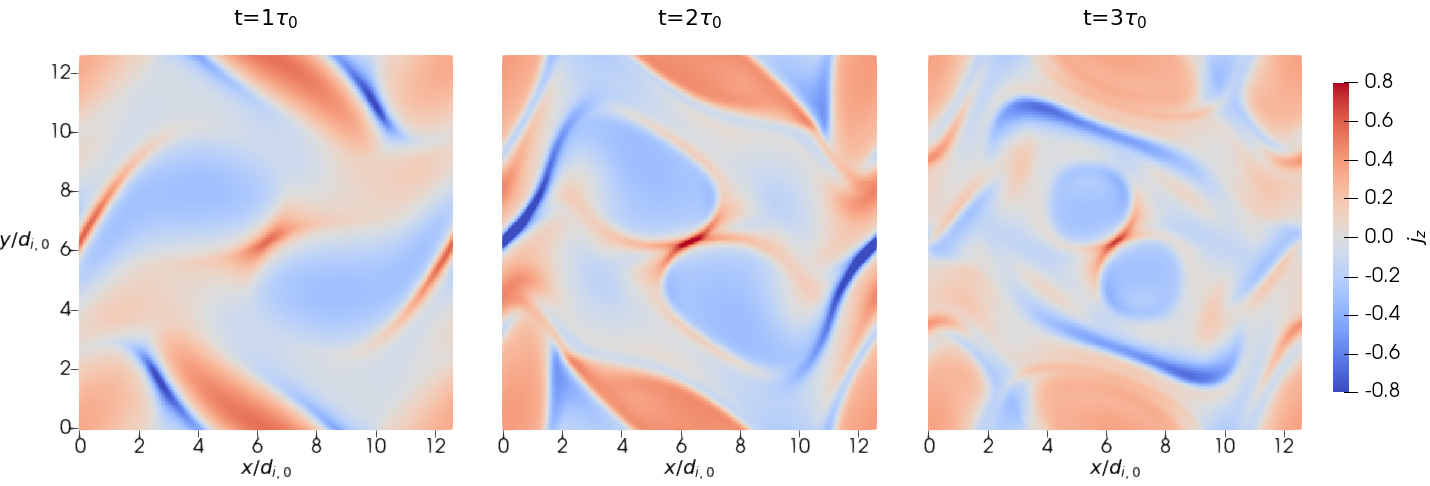}
\includegraphics[width = 0.9\textwidth]{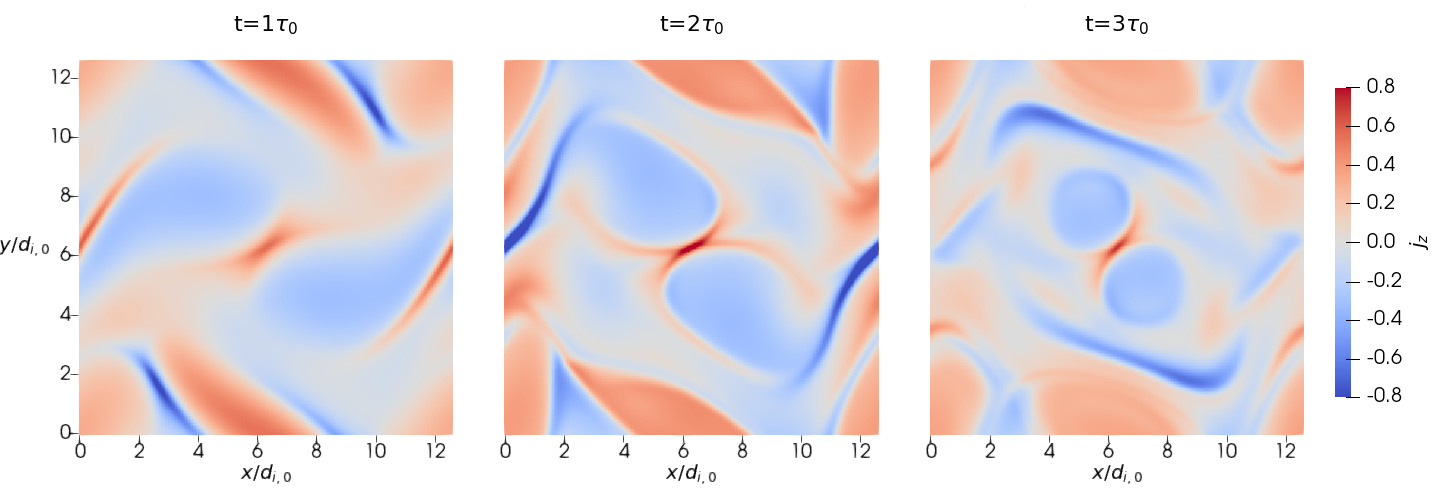}
\caption{Out-of-plane current density $j_z$ after one, two, and three large eddy turnover times $\tau_0=L/(2\pi \delta_u)$ using the AF scheme (upper panel) and the PFC scheme (lower panel). Note that all DoF are shown for the AF simulation.}
\label{fig:OT}
\end{figure}

In \autoref{fig:OT}, we compare the out-of-plane current density at one, two, and three large eddy turnover times between AF, resolved with $128^2\times20^3$ cells, and PFC, resolved with $256^2\times32^3$ cells. At these resolutions, the results are comparable, with AF having slightly sharper structures and a better preservation of the angularity of the vortex. The timestep was chosen according to $\nu=\text{max}(\mid v_e\mid)\Delta t / \Delta x = 0.1$. As AF uses half the number of cells in a spatial direction, the number of time steps required for this simulation was half the one needed to complete the PFC simulation.

\section{Concluding Discussion and Remarks}
In this paper, we present a scheme for solving the Vlasov--Maxwell system using the novel Active Flux (AF) method in combination with the classic Finite-Difference Time Domain solver for Maxwell's equations. Operator splitting is applied to reduce the dimensionality of the problem. This way, a one-dimensional version of AF can be utilized on (one-dimensional) fibers of the high-dimensional grid, which considerably reduces computational costs. AF evolves additional degrees of freedom per cell, from which the numerical fluxes are computed. This allows for a compact stencil both in space and time and results in a better resolution in configuration and velocity space compared to traditional Finite Volume schemes at the same number of cells. The Positive and Flux Conservative (PFC) method is used here as a state-of-the-art benchmark in our numerical experiments. The AF based scheme involves then the same number of independent degrees of freedom as the PFC based scheme does for twice the number of cells.
Owing to the compact stencil and the maximized CFL limit inherent to Active Flux methods, we achieve higher physical accuracy at lower compute time. The increased accuracy can mainly be attributed to lower dissipation resulting from the cell-local polynomial reconstruction, that also leads to a somewhat reduced anisotropy. We also point out that the violation of Gauss' law in regions of under-resolved steep gradients can quickly become detrimental to accuracy and stability of a scheme and is worth correcting. In this paper, we used a classical Boris correction term due to \cite{Birdsall}. Extending the approach that was proposed in \cite{crouseilles2014} for a semi-Lagrangian discretization or using hyperbolic/parabolic \textit{cleaning} strategies \cite{munz-etal:2000} could be more efficient and should be considered in future work.

\section*{Acknowledgements}

We gratefully acknowledge the Gauss Centre for Supercomputing e.V. (www.gauss-centre.eu) for funding this project by providing computing time
through the John von Neumann Institute for Computing (NIC) on the GCS Supercomputer JUWELS \cite{JUWELS} at Jülich Supercomputing Centre (JSC).
Computations were conducted on JUWELS/JUWELS-booster and on the DaVinci cluster at TP1 Plasma Research Department. R.G., G.G. and K.K. acknowledge funding from the German Science Foundation DFG through the research unit ``SNuBIC'' (DFG-FOR5409, project ids 463312734,  and 530709913). We are grateful for fruitful discussions with Christiane Helzel and Yanick Kiechle.

	\bibliography{bibliography}
	
\end{document}